\documentclass[twocolumn,secnumarabic,amssymb, nobibnotes,aps,prl,superscriptaddress]{revtex4-1}
\usepackage{graphicx}
\usepackage{textcomp}
\usepackage{amsmath}
\usepackage{commath}
\usepackage{color}
\usepackage{appendix}

\newcommand{\sg}[1]{}
\renewcommand{\sg}[1]{{\color{red}{#1}}} % Comment out to leave notes out (final version)

\begin{document}

\title{Bandgap tuning in kerfed metastrips under extreme deformation}
\author{Caleb Widstrand}
\affiliation{Department of Civil, Environmental, and Geo- Engineering, University of Minnesota, Minneapolis, MN 55455, USA}
\author{Negar Kalantar}
\affiliation{Architecture Division, California College of the Arts, San Francisco, CA 94107, USA}
\author{Stefano Gonella}
\affiliation{Department of Civil, Environmental, and Geo- Engineering, University of Minnesota, Minneapolis, MN 55455, USA}

\begin{abstract}
    \noindent The process of kerfing enables planar structures with the ability to undergo dramatic out-of-plane deformation in response to static loads. Starting from flat and stiff sheets, kerfing allows for the formation of a wide variety of unconventional free-form shapes, making the process especially attractive for architectural applications. In this work, we investigate numerically and experimentally the bandgap behavior of densely cut kerfed strips. Our study reveals a rich landscape of bandgaps that is predominantly ascribable to the activation of resonant sub-units within the kerf unit cells. We also document how the extreme deformability of the strips under twisting and bending loads, enhanced by the meandering cut pattern, can serve as a powerful bandgap tuning mechanism.
		\vspace{0.4cm}
\end{abstract}

	\maketitle

Kerfing (or relief cutting) is a perforation process consisting of introducing an intricate, periodic pattern of cuts, referred to as a \textit{kerf}, in a thin structure in a fashion aimed at deliberately increasing its compliance~\cite{Kalantar2018_CAADRIA,Mitov2018_NNJ}. Kerfing can be performed via laser or water-jet cutting on a variety of material platforms, including composites, such as medium-density fiberboard (MDF), and metals like aluminum or steel. The spatial periodicity makes the design of kerf patterns easily achievable via automatic generation algorithms~\cite{Zarrinmehr2017_CG}.

By removing solid material and leaving behind a network of interconnected slender elements, kerfing endows an initially flat, homogeneous and relatively stiff sheet with an excess of compliance, making it capable of undergoing dramatic out-of-plane deformation in response to static loads~\cite{Chen2020_ActaMech}. It is possible to classify kerf patterns according to three archetypal genealogies~\cite{Kalama2020}: 1) linear-cut patterns, involving only parallel cuts and leading to single curvature; 2) offset patterns, in which straight cuts are offset and connected at given angles, allowing double curvature; 3) meandering patterns (the focus of this Letter), featuring tightly interlocking chiral cuts realizable with a variety of cut densities, also conducive to double curvature. The small-scale heterogeneity of densely cut meandering kerfs endows them with extreme deformability that enables morphing into free-form structures. Deformations typically occur within the elastic limit of the material, and are fully reversible upon unloading although, working with composites, it is possible to lock the deformed shapes using epoxy resins. %Alternatively, permanently deformed shapes have been achieved with metal sheets, deforming their elements beyond the plastic limit. 
With respect to their ability to undergo large deformation as a result of strategically placed cuts, kerfed sheets are conceptually similar to kirigami structures~\cite{Zhang2015_PNAS,Tang2019_PNAS} and structures in which out-of-plane instabilities are promoted by a cut pattern~\cite{Cho2014_PNAS,Celli2018_SM}.

Kerfed panels are especially appealing for architectural applications, where free-form structures have become increasingly ubiquitous for the design of building facades~\cite{Pottmann2008_ACM,Andrade2017_FoAR} thanks to their ability to meet simultaneously aesthetic and functional requirements. In this context, kerfed sheets represent a light-weight material that naturally lends itself to being reconfigured in free-form shapes of diverse, pre-designed complexity. An additional benefit lies in the fact that kerfed sheets can be shipped flat and easily deformed on-site before installation, with major implications for transportation and logistics. %The perforated nature and the inherent reconfigurability of kerfs may also lead to interesting air flow and light management, which are important design factors for the built environment. \sg{[REF.s]} 

%\begin{figure}[b]%[!htpb]
%	\centering
%	\includegraphics[width=0.48\textwidth]{MDF_Strips.pdf}
%	\caption{Extreme deformability of kerfed strips. (a) Bending with sharp flexural hinges. (b) Deformation into  M$\ddot{\textrm{o}}$bius strip. (c) Twisting of bi-kerfed strip (dense $+$ coarse cuts) with torsional hinges. (d) Deformation into helical structure. (e) Schematic showing potential of kerfed strips with tunable twist for facades with programmable opening patterns.} \label{MDF_Strips}
%\end{figure}

The periodic microstructure of kerfs allows classifying them as examples of elastic phononic crystals (PCs) and metamaterials~\cite{Kushawaha1993_PRL,Sigalas1993_SSC,Jensen2003_JSV,Khelif2006_PhysRevE,Spadoni2009_WM}, as well as a special subfamily of perforated plates~\cite{Slot1971_JEI,Jing2000_AIAA,Pedersen2004_SMO,Hedayatrasa_2016_JMPS}. We therefore expect them to exhibit a whole array of dynamical properties classically observed in PCs, and specifically the ability to open bandgaps~\cite{Liu2000_Science,Martinsson2003_QJMAM,Gonella2008_JSS,Hsu2010_APL,Wang2015_PhyRevB}. 
The first objective of this study is to assess the effectiveness of kerfed structures as bandgap materials. % capable of filtering mechanical vibrations with given frequency contents. 
%With reference to architectural applications, the bandgap opening capability would endow kerfed sheets with the ability to manage, filter and attenuate mechanical vibrations and sound, endowing them with an additional capability to manage wind loads or seismic events.
Recently, significant interest has revolved around the design of tunable structures capable of adjusting their behavior in response to externally applied stimuli. % to adapt to evolving operating conditions. 
A number of strategies have been proposed, in which the tuning is brought about via electro-mechanical~\cite{Casadei2012_JAP} or magneto-mechanical~\cite{Bilal2017_AdvM,Schaeffer2015_JAP} coupling, or by deformation or buckling~\cite{WangCasadei2014_PRL}. In this vein, the second objective of this study is to assess whether the intrinsic extreme deformability of kerfs can be leveraged as an effective tuning mechanism. %The remainder of this Letter is devoted to addressing these two objectives. % through a series of carefully designed experiments.

\begin{figure*}[!htpb]
	\centering
	\includegraphics[width=1\textwidth]{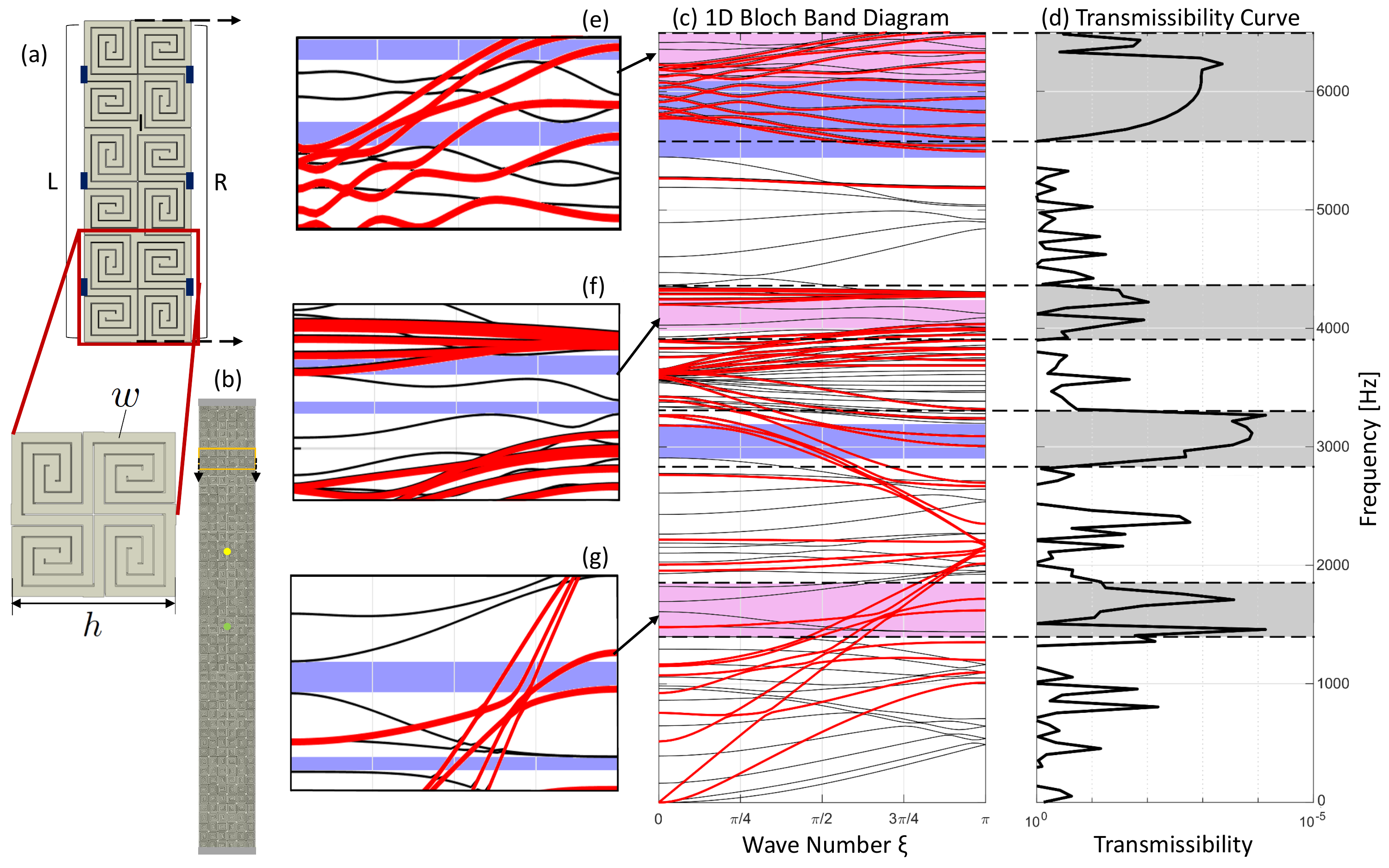}
	\caption{(a) Macro-cell of densely cut kerfed strip. (b) 3$\times$24 strip with excitation and response sampling points highlighted in green and yellow, respectively. (c) Macro-cell band diagram with main flexural bandgaps highlighted in purple and secondary regions of attenuation in bright pink. (d) Transmissibility curve from harmonic analysis of finite strip. (e-g) Details of secondary attenuation regions revealing clusters of narrow bandgaps separated by nearly-flat modes} \label{Numerics}
\end{figure*}

Here we focus our attention on kerfed strips, obtained by tessellating in one dimension a macro-cell consisting of a finite number of kerf cells. This choice is motivated by the extreme deformability and reconfigurability that are achievable working with strips, even under relatively small loads. The deformability of the strips results from the combined effect of the compliance inherently provided by their slenderness and the geometric softening introduced by the kerfing process. %This characteristic makes them an agile experimental material platform amenable to varieties of table-top experiments. 
%Examples of extreme deformations achievable in kerfed MDF strips are shown in Fig.~\ref{MDF_Strips}(a-d). Another source of inspiration comes again from the realm of architecture. We envision the design of facades consisting of arrays of kerfed strips that can be twisted, either in bulk or in subsets, according to an variety of twist rates, to obtain a plethora of openings patterns that can be tuned on demand. While actual architectural implementations of this concept are a long-term proposition, they offer an intriguing opportunity for the application of this technology to light, air circulation and sound management problems. 

Let us model a strip as shown in Fig.~\ref{Numerics}(b) via 1D tessellation of the three-cell-wide macro-cell with a densely cut kerfing pattern shown in Fig.~\ref{Numerics}(a). % in the direction of the black arrows. 
The geometry of the kerf is as follows: cell size $h=1.905 \, \textrm{cm}$, ligament width $w=1.143 \, \textrm{mm}$ and out-of-plane strip thickness $0.762 \, \textrm{mm}$. We assume the material properties of 304 stainless steel: Young's modulus $E = 193 \, \textrm{GPa}$, Poisson's ratio $\nu$ = 0.27 and density $\rho$ = 7870 $\mathrm{kg/m^3}$. We build a finite element (FE) model of the macro-cell, using a mesh of 8-node isoparametric elements featuring two elements through the thickness. This modeling process is done using the software GMSH~\cite{Geuzaine_GMSH}. %A discussion of the mesh refinement process conducted to ensure reliability of the FE results are given in the SM. 
To obtain a band diagram, %for the strip,
we follow the canonical steps of 1D Bloch analysis, %(main details recalled in the SM), 
whereby we apply 1D Bloch conditions between the $\textrm{L}$ and $\textrm{R}$ edge nodes, here confined to the protruding elements marked in blue, leaving the other edges free. %It is interesting to note that, because of the cuts and our selection of unit cell, the connections between the macro-cell and its neighbors is confined to the protruding elements marked in blue.
We then sweep the scalar non-dimensional wavenumber $\xi$ in the irreducible Brillouin zone, and, for each $\xi$, we solve an eigenvalue problem involving the reduced stiffness and mass matrices of the macro-cell, yielding the %branches of the 
macro-cell dispersion branches. The end result is the band diagram of Fig.~\ref{Numerics}(c), whose high modal density stems from the ability of the 3D FE model to capture all wave polarizations. Through a comparison of the branches against those obtained from a companion planar model of the macro-cell based on 2D elasticity %and a mesh of 4-node elements
(details in the SM), we can discriminate the strictly in-plane modes, marked in bold red, from those that possess significant out-of-plane character, shown in black. While no total bandgaps are found in the frequency interval of interest, we recognize two flexural bandgaps, marked in purple, in which flexural modes are evanescent but in-plane waves are allowed to propagate.
% WE CNA ADD IT IF NEEDED
%%%%%%%%%%%%%%%%%%%%%%%%%%%%%%%%%%%%%%%%%%%%%%
%In the SM, we show for comparison the band diagram obtained subjecting a unit cell to 2D Bloch analysis, which captures bulk modes of a two-dimensional kerfed sheet with the same cut pattern and density. Overall, the macro-cell band diagram is characterized by a higher modal density, due to the emergence of strip modes featuring localization at the free edges (the long sides fo the strip) and therefore uniquely germane to the strip. Nevertheless, both diagrams show good agreement in terms of onset and width of the flexural bandgaps.
%%%%%%%%%%%%%%%%%%%%%%%%%%%%%%%%%%%%%%%%%%%%%%
\begin{figure}[t]
	\centering
	\includegraphics[width=0.5\textwidth]{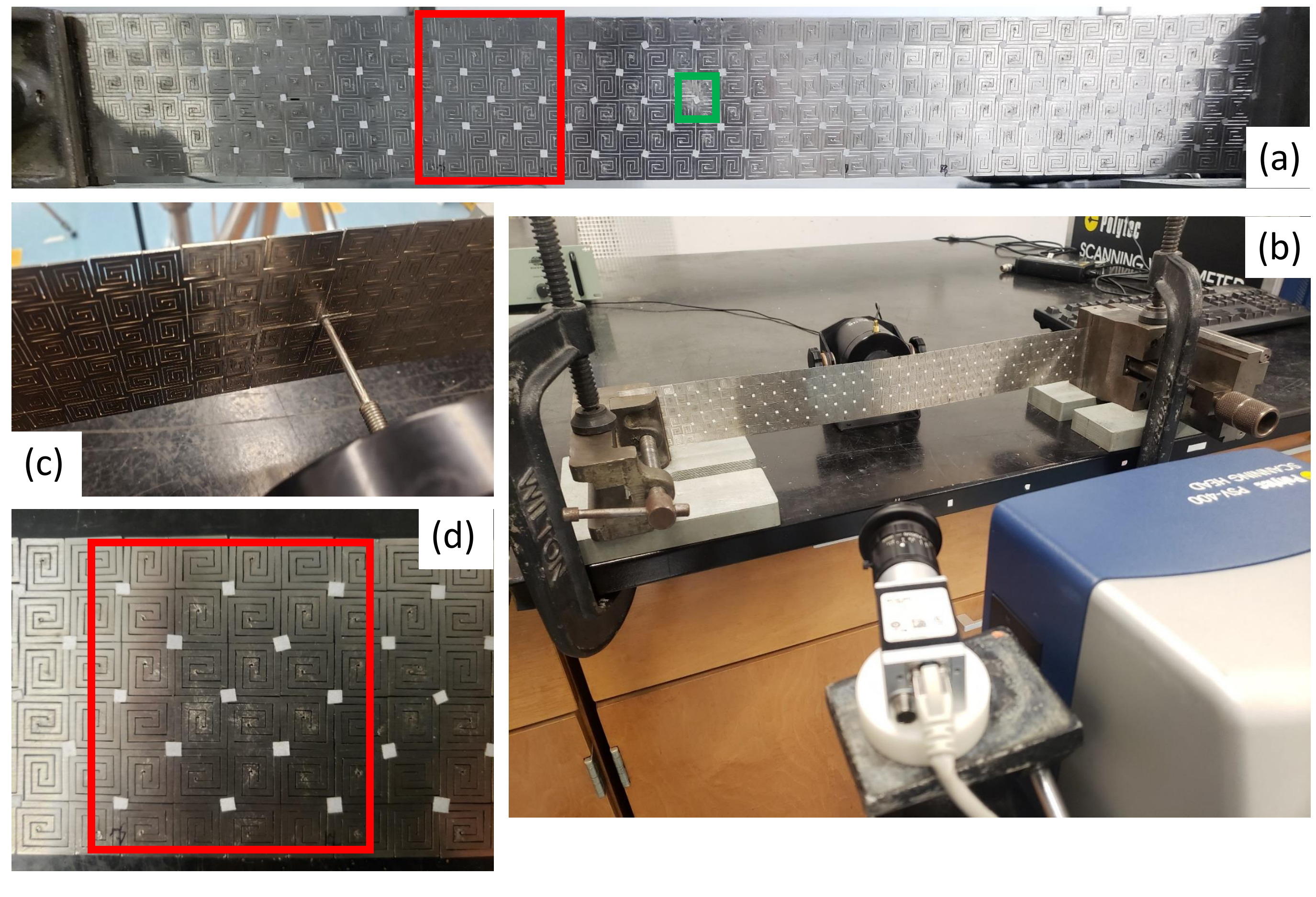}
	\caption{(a) Flat densely cut kerfed strip with sampling points highlighted in red and excitation point marked in green. (b) Experimental setup showing clamped specimen and laser vibrometer in the foreground. (c) Rear face of specimen showing placement of the shaker. (d) Close-up of sampling region.} \label{Setup}
\end{figure}
% between this band diagram and the band diagram produced from 2D Bloch analysis on a single kerfed unit cell, and analysis of the mode shapes at the onsets of these bandgaps revealed the activation of flexural resonators interior to the unit cell introduced by the meandering cutting pattern as the mechanisms responsible for the bandgap behavior (see supplemental materials). 

We complement Bloch analysis with a full-scale steady-state analysis on the 3$\times$24 kerfed strip depicted in Fig.~\ref{Numerics}(b). We prescribe fixed boundary conditions at both ends and we apply a sustained harmonic excitation at the mid point (green dot), sweeping the frequency. The displacement is sampled at the point marked by the yellow dot %located a few macro-cells away from the excitation point, 
and normalized by the displacement of the excitation point to construct a curve of transmissibility vs. frequency, plotted in Fig.~\ref{Numerics}(d). The curve features several regions of attenuation, highlighted in gray. %, where the transmissibility drops by several orders of magnitude.
It is clear that only two of these regions can be pinpointed to the major flexural bandgaps discussed above. To elucidate the sources of the other attenuation zones, we zoom in on the corresponding intervals of the band diagram marked in bright pink, detailed in the Fig.~\ref{Numerics}(e-g) insets. We observe the appearance of clusters of narrower bandgaps, interspersed by nearly flat modes characterized by slow propagation velocities, which, in concert, make these intervals globally unfavorable for transmission.
% MAYBE ADD THIS
%We believe that the discrepancy between the band diagram and the transmissibility curve at the onset of the upper bandgap is in part caused by a mismatch between the mesh refinements that we could afford in the two analyses. The mesh size required to discretize the finite strip results in massive system matrices that pushed us the limit of our computing capabilities. This forced us to work with a slightly coarser mesh than the one used for the macro-cell. The differences between the two meshes are negligible at lower frequencies, but become noticeable in the range of the above-mentioned gap. These considerations suggest that a superior agreement than what showed between Figs.~\ref{Numerics}b and c, should be achievable even at high frequencies using a more refined computational model.

From the numerics, we have evinced a rich bandgap landscape and an even richer availability of transmission attenuation intervals, which suggests that densely cut kerfed strips work efficiently as 1D filters. We now seek experimental validation of these findings through a test whose key features are summarized in Fig.~\ref{Setup}. The strip shown in Fig.~\ref{Setup}(a) is kerfed from a steel plate according to the geometry used for the model. Fig.~\ref{Setup}(b) shows the strip ends clamped using a pair of vices and the scanning head of a laser vibrometer (Polytec PSV-400M-3D, used in 1D mode). The excitation is prescribed out-of-plane with a shaker at the point boxed in green on the rear face of the strip (Fig.~\ref{Setup}(c)). The red box denotes the response sampling region; in the Fig.~\ref{Setup}(d) detail one can appreciate patches of retro-reflective tape applied at the scan points, conveniently located at the connections between cells where uncut areas are more amenable to scanning. %, mimicking the numerical simulation conducted previously on the full-scale kerfed strip model.

\begin{figure*}[!htpb]
	\centering
	\includegraphics[width=1\textwidth]{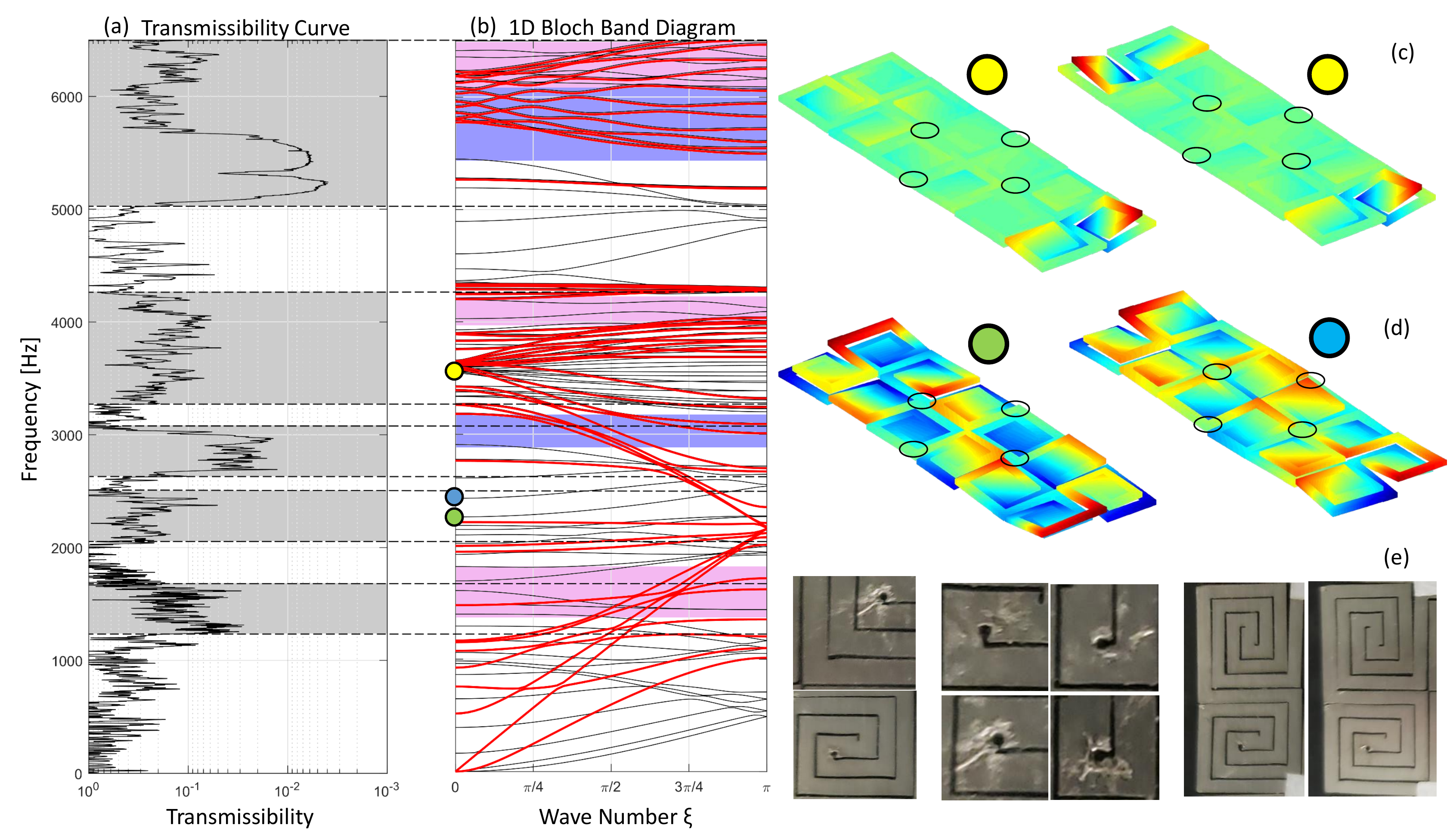}
	\caption{(a) Experimental transmissibility curve with regions of attenuation highlighted in gray. (b) 1D Bloch band diagram juxtaposed for reference. (c)-(d) Mode shapes corresponding to experimentally observed frequency ranges of attenuation, with color map corresponding to normalized out-of-plane displacement. Points where the strip is sampled in the experiments, circled in black, featuring negligible displacement. (e) Close-ups of specimen cells, revealing defects and non-idealities.} \label{Experiment} % causing inherent and unpredictable differences between the specimen and the numerical model.} 
\end{figure*}

We prescribe a broadband pseudo-random excitation %at the aforementioned excitation point using a shaker device attached to the reverse side of the specimen as shown in Fig.~\ref{Setup}(c). In this way, we can 
to excite a broad range of frequencies. % simultaneously in the specimen. %, thus allowing us to study its broadband frequency response. 
A drawback of using this kind of signal is that the shaker output energy spectrum decays sensibly with frequency in the interval of interest, thus failing at supplying enough energy at high frequencies. To overcome this, we apply the signal in 2 kHz increments between 0 and 8 kHz, prescribing higher amplitudes in the upper half of the range, %between 4 and 8 kHz,
such that a roughly equal amount of energy is injected into the system for all frequencies. %Within each interval, 
We measure the out-of-plane velocity at the designated sampling points, we average them and finally normalize this average by the response of the excitation point to construct a measure of velocity transmissibility, plotted against frequency in Fig.~\ref{Experiment}(a). %The interval-wise data is stitched together to yield the curve of transmissibility vs. frequency of Fig.~\ref{Experiment}(a).%, a measure of the input/output frequency response of the specimen at several frequency values. 

 % ranging between -1 (blue) and 1 (red)
 
%We can visualize the response of the kerfed strip as a transmissibility curve in Fig.~\ref{Experiment}(a) which plot the transmissibility through the specimen against frequency from 0 to 6500 Hz. The transmissibility is a fraction of 1, and significantly lower values would therefore correspond to attenuation of the response of the specimen; 
We highlight in gray all the regions of attenuation. % observed in the interval of interest. 
Comparing against the band diagram juxtaposed in Fig.~\ref{Experiment}(b) for reference, we observe an overall satisfactory agreement between these intervals and the regions of attenuation predicted by Bloch analysis. %, whether due to major flexural bandgaps or to clusters of smaller bandgaps.
Specifically, we can attribute the large drops in transmissibility starting around 2.6 $\textrm{kHz}$ and 5 $\textrm{kHz}$ to the two major flexural bandgaps highlighted in purple, albeit with some discrepancies in the onset frequencies that we will discuss below. Additionally, we can pinpoint the attenuation zone starting around 1.2 $\textrm{kHz}$ and the upper portion of the one starting at 3.2 $\textrm{kHz}$ to the narrow bandgap clusters previously identified in those intervals. 

However, the experimental data displays other dips in transmissibility that cannot be attributed to any obvious attenuation mechanism indicated by the band diagram. Some rationale into this discrepancy can be gained by looking at the mode shapes associated with selected spectral points in these ranges. %We should note that the mode shapes show us the deflection shapes of the unit cell and they are computed using the eigenvectors from the eigenvalue problem solved in Bloch analysis. 
For example, let us consider the two mode shapes at the spectral point highlighted by the yellow dot at 3.6 $\textrm{kHz}$, plotted in Fig.~\ref{Experiment}(c) with color map proportional to the normalized out-of-plane displacement. We observe a localization of deflection at the free edges of the macro-cell (the long sides of the strip), deeming these as waveguide modes germane to the 1D geometry of the strip, while the bulk of the strip experiences negligible motion. The black circles superimposed to the figure mark the location of the scan points where we sample the response. The decision to cluster the sampling in the interior of the strip is dictated by practical difficulties encountered in reliably scanning the peripheral portion of the specimen, mostly due to the abnormal lateral deformation and warping exhibited by the edge cells as the result of the cutting process. We recognize that, at this frequency, our scan point selection cannot pick up any response associated with these modes, thus outputting a spurious attenuation zone. Similar considerations can be made about the mode shapes for the spectral points marked by the green and blue dots plotted in Fig.~\ref{Experiment}(d). While these are definitely bulk modes, %that do travel along the axis of the strip, 
they happen to feature nodes (points experiencing negligible displacements) at the scan point locations, which makes them transparent to our sensing strategy. Here, a denser scan would likely detect this modal response and absorb the apparent bandgap conditions, but such refinement would be prohibitive to achieve with a densely cut kerfed structure. This analysis offers a stark reminder that even experiments conducted with sophisticated measurement tools can detect a variety of attenuation mechanisms that are spurious byproducts of the sensing strategy. % and/or of the specimen non-idealities. %It confirms that a full and reliable dynamical analysis shall always involve a side-by-side mutually-informing inspection of numerical and experimental data.
%and we find these exhibit a similar behavior to the aforementioned modeshapes where these is little deflection at points connecting unit cells. We conclude that for some frequency ranges, even those where we do not observe bandgaps in the overall band structure, we are afforded attenuation in the frequency response of the kerfed strip due to a choice of sampling that is not amenable to these particular frequency ranges.

\begin{figure*}[t]
	\centering
	\includegraphics[width=1\textwidth]{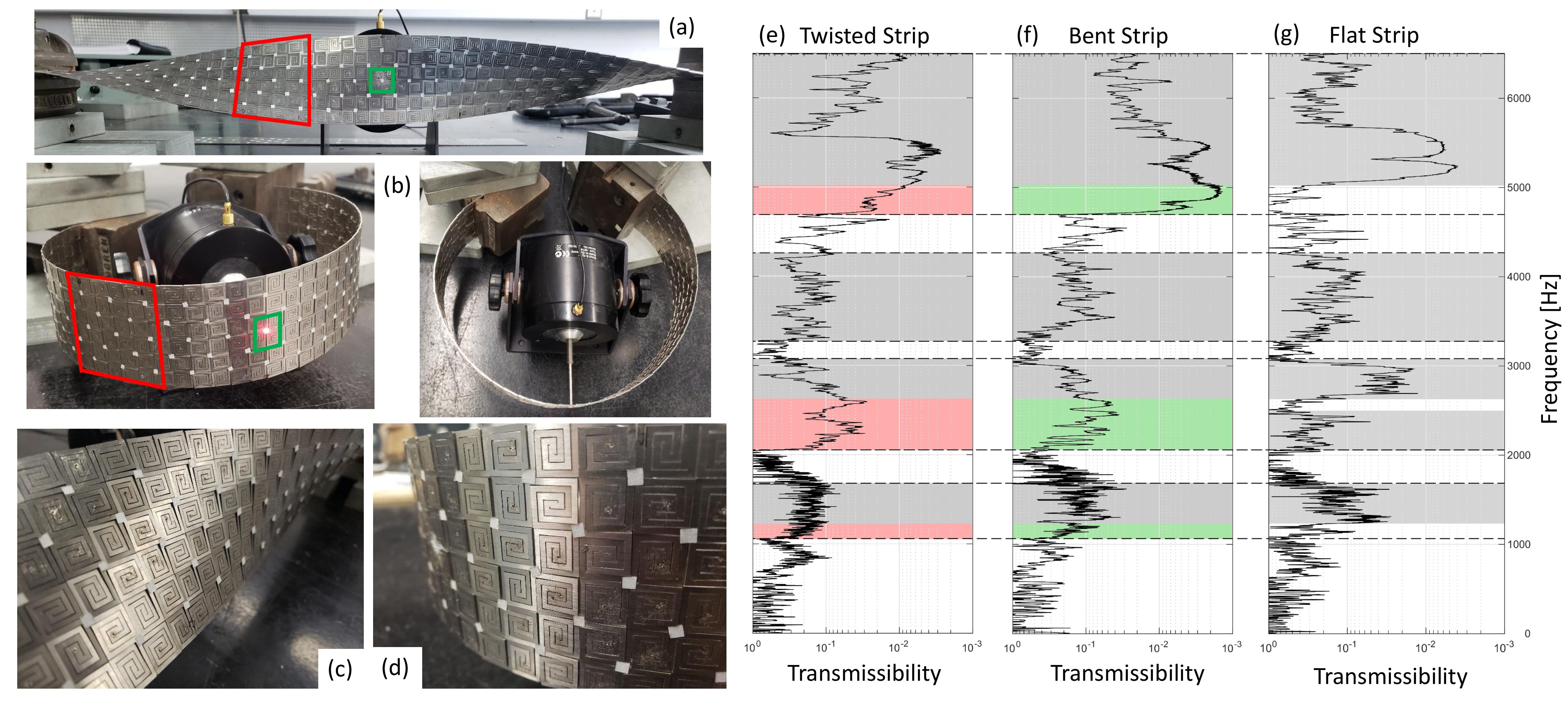}
	\caption{(a) 180$^{\circ}$ twisted strip. (b) Bent strip forming nearly closed loop. (c-d) Close-ups of out-of-plane popping mechanisms enabled by the twisting and bending deformations, respectively. (e) Experimental transmissibility curve for twisted strip. Pink regions denote changes in the attenuation zones brought about by twisting. (f) Transmissibility curve for the bent strip, with green regions denoting the corrections induced by bending. (g) Flat strip transmissibility curve.} \label{Tuning}
\end{figure*}

We also report other discrepancies that appear consistently across experimental takes and that cannot be as easily explained by the morphology of the mode shapes or by the sampling procedure. For instance, we observe some frequency downshifts in the experimental data at the onset of both major flexural bandgaps. We believe that the root of these discrepancies lies in a series of defects or geometrical non-idealities embedded in the specimen during the fabrication process, shown in Fig.~\ref{Experiment}(e), which cause the specimen to deviate from the model.
For example, we detect a ``waviness" of the cut path that causes a variation in the width of the beams along the kerf pattern. Additionally, we note the presence of fillets where the cutting path terminates, which locally reduce the beams width. % and soften the response. % of the strip. %, possibly contributing to an overall lowering of the frequency of the modal branches.
Finally, we observe some residual in-plane deformation of the cells located near the edges of the strip, resulting from the lateral pressure exerted during the cutting process. % and exacerbated by the thinness of the kerf ligaments. 
The random distribution of such defects over the specimen domain %is largely random and ultimately 
inevitably relaxes the perfect periodicity assumptions underpinning the numerical model. %However, we want to emphasize that the transmissibility curve exhibits key features, the major bandgaps and collections of smaller bandgaps, that we can directly attribute to the band structure computed via Bloch analysis.

Having established the attenuation behavior of a flat strip, we investigate the possibility to tune its bandgap landscape by subjecting it to a variety of large deformation modes. % enabled by their kerf-induced compliance. 
In our first scenario, we consider a strip twisted by 180$^{\circ}$ from end-to-end, as shown in Fig.~\ref{Tuning}(a). In the second case, we bend the strip to form a nearly closed loop, as shown in Fig.~\ref{Tuning}(b). We note that, in both cases, the applied deformation is enabled by a dramatic reconfiguration of the cell's internal features, which experience localized out-of-plane deflections, effectively ``popping" out of the strip plane as shown in Figs.~\ref{Tuning}(c) and (d). As visible in Figs.~\ref{Tuning}(a) and (b), the deformation introduces a relative rotation between the sampling region and the excitation region. %, exposing them differently to the laser head.
This warrants a slight modification in our testing protocol, whereby we need to reorient the laser head in-between scans of the two regions to ensure the we always take measurements in a direction locally perpendicular to the strip. %Otherwise, we employ the same settings and procedures as with the flat strip using 2 kHz intervals of pseudorandom excitations.

We first examine the transmissibility for the twisted case shown in Fig.~\ref{Tuning}(e). We observe that the attenuation zones, while overall matching those of the flat strip highlighted in gray and recalled in Fig.~\ref{Tuning}(g), differ in both shape and width. Specifically, the lowest and highest bandgap openings (around 1 and 5 \textrm{KHz}) display a lower onset compared to their flat strip counterparts, which de facto stretches the intervals of attenuation by the amount marked in pink. Additionally, we observe a merging of the attenuation zones between 2 and 3 \textrm{KHz}, %(of which one was associated to a major flexural bandgap and that other was traceable to the experimentally unfavorable mode shapes available in that interval), 
which comes at the expense of the major bandgap depth. 
Turning our attention to the bending case of Fig.~\ref{Tuning}(f), we observe a remarkably similar landscape of modifications in the transmissibility curve, here shaded in green. While the available data does not allow reaching a definitive explanation for these corrections, or for their lack of sensitivity to the type of deformation, we propose the following line of interpretation. As the strip deforms, the ``pop-out" mechanisms observable in Figs.~\ref{Tuning}(c) and (d) modify the internal geometry of the cells, thus tuning their resonant characteristics. Incidentally, these geometric modifications are similar between twisting and bending. Since we were able to pinpoint several bandgap formation mechanisms to the resonant behavior of the internal microstructure, it is reasonable to expect these morphological changes in the resonators to affect the onset of bandgaps. Interestingly, the bandgap widening observed here is qualitatively reminiscent of that observed in metamaterials upon randomization of their resonant microstructure~\cite{Celli2015_APL}. The fact that the cell reconfiguration induced by deformation is extremely non-uniform over the strip hints at randomness as a possible co-factor causing these corrections.

In conclusion, we have demonstrated experimentally that kerfed metastrips feature a rich filtering behavior associated with the activation of resonant mechanisms in the kerf cells. This behavior can be modified through the application of extreme twisting and bending deformation, indicating that the compliance induced by kerfing can be leveraged as an effective bandgap tuning mechanism.

%In conclusion, we have demonstrated experimentally that kerf metastrips feature a rich bandgap filtering behavior that is tunable upon application of extreme twisting and bending deformation. While more work is necessary to pinpoint these corrections to specific mechanisms, the strengths of these effects allows to conclude that the compliance induced by kerfing can be used as an effective tuning mechanism for the dynamics of one-dimensional metastructures.

The authors acknowledge support from the National Science Foundation (EAGER grant CMMI - 1911678). N.G. also acknowledges partial support from NSF grant CMMI - 1913688.  We are grateful to A. Muliana at Texas A$\&$M 
for useful discussions.

%This work is funded by the National Science Foundation (EAGER Grant CMMI - 1911678 and CMMI - 1913688). We are grateful to A. Muliana's group %at Texas A$\&$M 
%for initial support with specimen fabrication.

\bibliographystyle{unsrt}
\bibliography{Metastrips_Bibliography}

\onecolumngrid

\newpage

\section*{\Large{SUPPLEMENTAL MATERIAL}}

\subsection*{\Large{Analogies and differences between 1D and 2D Bloch analysis}}

\begin{figure*}[!htpb]
	\centering
	\includegraphics[width=1\textwidth]{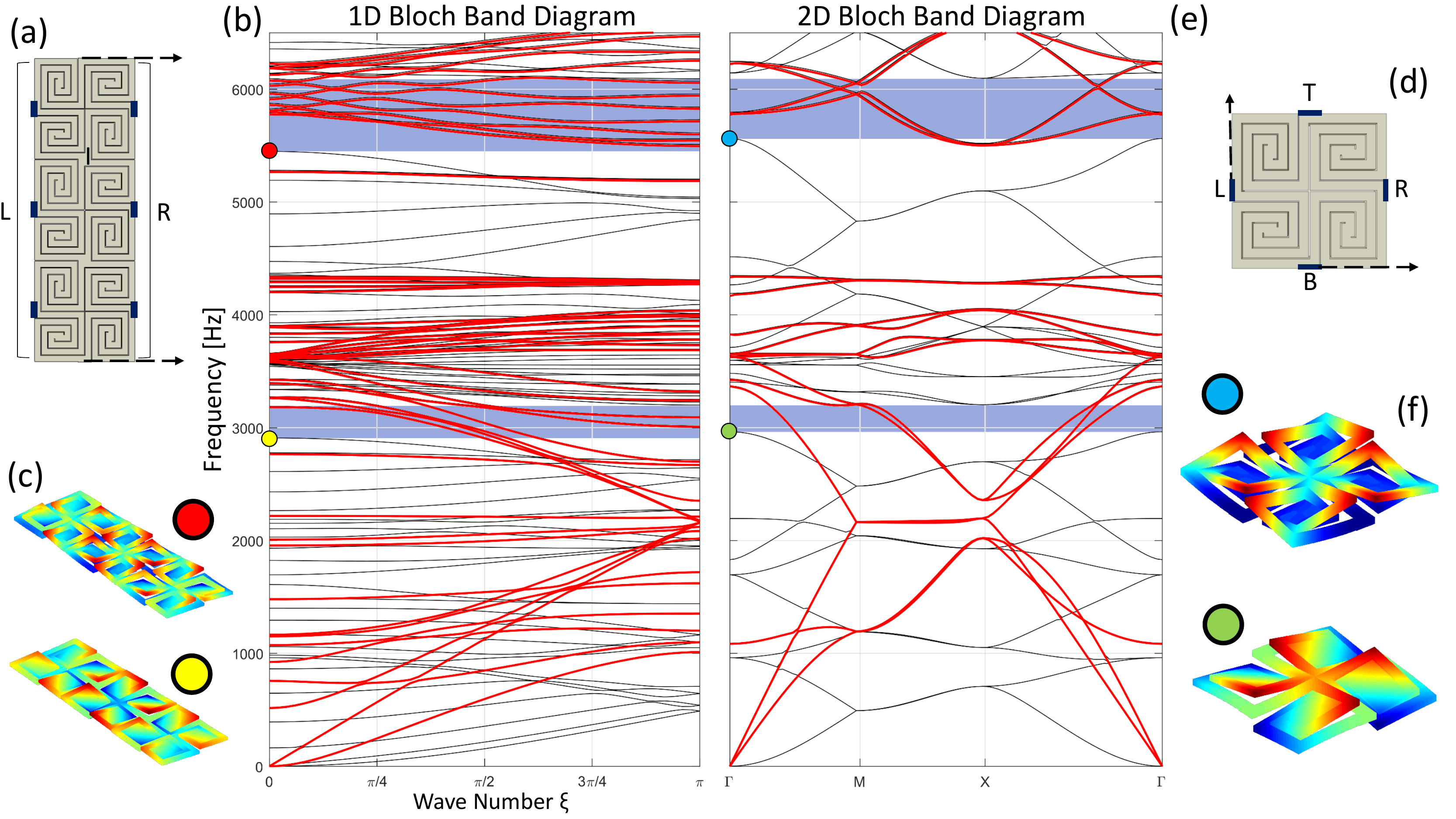}
	\caption{(a) Dense unit macro-cell of 1D Bloch analysis. (b) Band diagram computed from 1D Bloch analysis with major flexural bandgaps in highlighted purple. (c) Mode shapes of the macro-cell at the onsets of the flexural bangaps, featuring activation of flexural resonators in the interior of the unit cell. (d) Unit cell of 2D Bloch analysis. (e) Band diagram computed from 2D Bloch analysis with flexural bandgaps highlighted purple agreeing reasonably well with those predicted by 1D Bloch analysis. (f) Mode shapes plotted at the onset of each bandgap, featuring similar activation of the cell interior flexural resonators as observed from the mode shapes of 1D Bloch.} \label{1D2DBloch_SM}
\end{figure*}

In order to emphasize the need to perform 1D Bloch analysis to properly capture the strip dynamics, we also perform a 2D Bloch analysis on a single densely kerfed unit cell with the same geometric and material characteristics, and we compare the results in Fig.~\ref{1D2DBloch_SM}.

We show, for reference, the macro-cell and the corresponding band diagram for the 1D Bloch analysis in Fig.~\ref{1D2DBloch_SM}(a) and Fig.~\ref{1D2DBloch_SM}(b). The 2D Bloch band diagram shown in Fig.~\ref{1D2DBloch_SM}(e) is computed from Bloch analysis on the densely kerfed unit cell in Fig.~\ref{1D2DBloch_SM}(d), upon application of Bloch conditions relating the left and right as well as the bottom and top sets of nodes. We observe that the 1D Bloch band diagram features several additional modes compared to its 2D counterpart, which correspond to waveguide modes germane to the strip, in which the deformation is localized along the strip long edges, that cannot be captured by the unit cell analysis. 

We find the availability of major flexural bandgaps, highlighted in purple in Fig.~\ref{1D2DBloch_SM} to be consistent between the two analyses in terms of both location and width. In the main body of the work, we also observed the availability of smaller bandgaps in the band structure of the macro-cell. These are not present in the 2D Bloch band diagram, hinting at the fact that their opening is related to mechanisms that are uniquely germane to the strip dynamics.

We also want to draw attention to the mode shapes at the onsets of the major flexural bandgaps, which we show for each analysis in Fig.~\ref{1D2DBloch_SM}(c) and Fig.~\ref{1D2DBloch_SM}(f), that exhibit strong flexural activation manifesting as the ``flapping" of units interior to the kerf pattern. Incidentally, the same type of internal flexural mechanism is consistently identified in both the 1D and 2D Bloch analysis. From these observations, we attribute the opening of the major bandgaps to the activation of resonant behavior in the microstructure of the kerfing pattern. However, it is important to point out that these are not locally resonant mechanisms as conventionally observed in lattices with internal microstructures \cite{Liu2000_Science}, in which the resonating elements are typically auxiliary components
that one can add or remove from the cell without altering the structure's static
stiffness or load-bearing capabilities. Here, the elements that resonate are an integral part of the connective pathways along with the solid portion of the plate that is retained during the cutting process. In a similar vein to the observed behavior of these kerfed structures, analysis of the dynamic response of periodic lattices of beams has shown that resonant bandgaps can be opened by resonance of the connective beams of the lattice~\cite{Wang2015_PhyRevB}.

\subsection*{\Large{Comparison between dense and coarse cut kerfed cells}}

We also investigate an alternative meandering cutting pattern featuring less turns in the cutting path, leading to overall widening of the beam elements and a stiffer response. We call this pattern the coarse cut, and we call unit cells with this particular pattern coarsely kerfed or coarse cells.
\begin{figure*}[!htpb]
	\centering
	\includegraphics[width=1\textwidth]{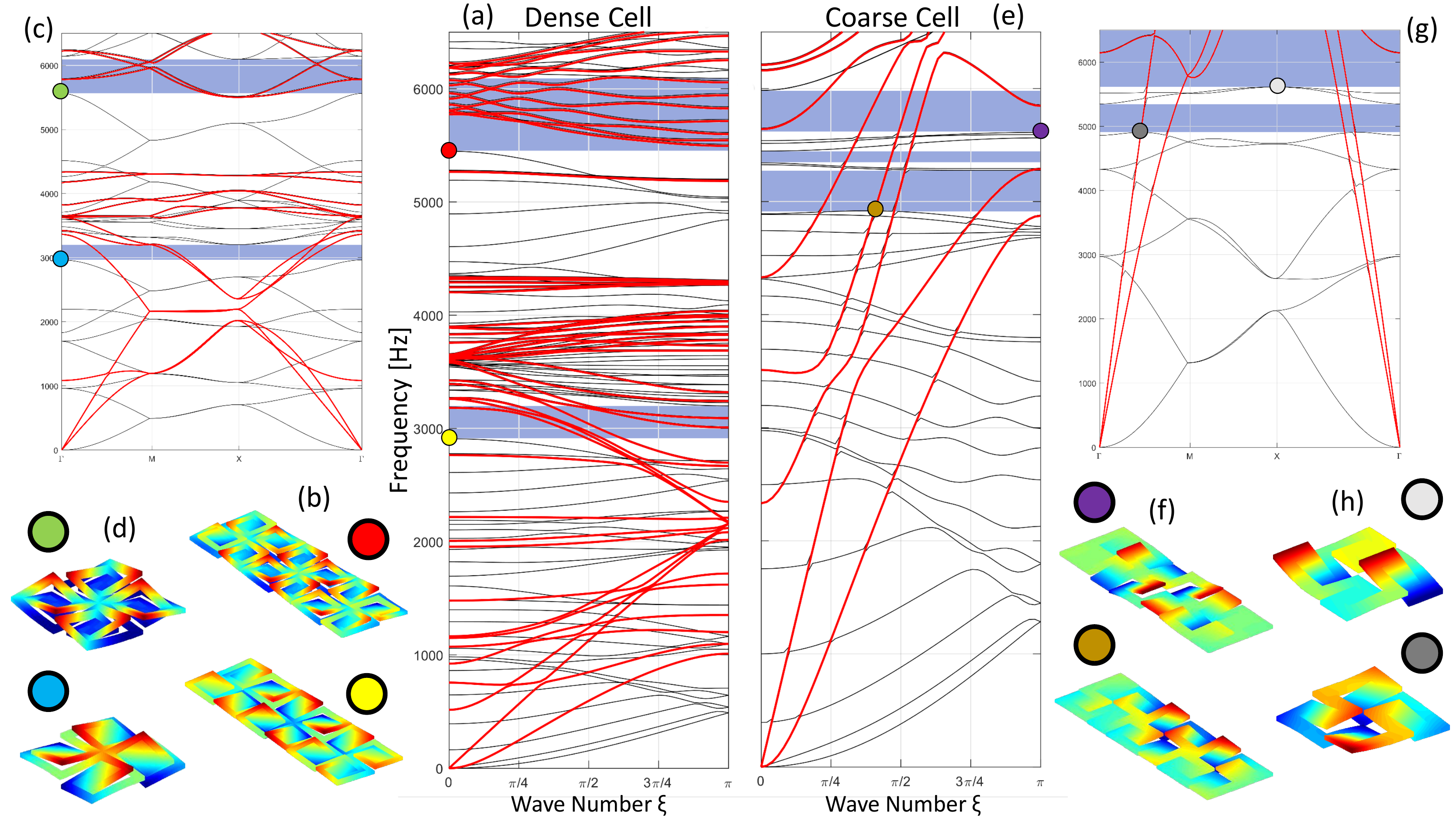}
	\caption{(a) 1D Bloch band diagram of dense cell. (b) Mode shapes at onsets of bandgaps. (c) 2D Bloch band diagram of dense cell. (d) Mode shapes at bandgap onset from 2D Bloch. (e) 1D Bloch band diagram of coarse cell featuring flexural bandgaps in purple. (f) Mode shapes at onset of the bandgaps demonstrating strong activation of flexural resonators. (g) 2D Bloch band diagram of coarse unit cell featuring bandgaps in a similar range as 1D Bloch. (h) 2D Bloch mode shapes plotted at bandgap opening that demonstrate the same flexural resonant behavior as the macro-cell.} \label{CoarseDenseCompare_SM}
\end{figure*}

Figs.~\ref{CoarseDenseCompare_SM}(a-d) feature the now familiar 1D and 2D Bloch analyses as well as mode shapes at the onset of the major bandgaps. Fig.~\ref{CoarseDenseCompare_SM}(e) shows the band diagram computed from 1D Bloch analysis on a coarsely kerfed macro-cell. Similar to the band diagram of the dense macro-cell, we identify major flexural bandgaps, highlighted in purple. However, we note that the bandgaps of the coarse cell are only clustered at high frequencies, implying that, with this configuration, we are only afforded bandgap behavior in a specific range of our frequency interval of investigation.

We compare the band diagram of the coarse macro-cell to the band diagram of a single coarse unit cell, shown in Fig.~\ref{CoarseDenseCompare_SM}(g), and we observe good agreement in the location and range of the bandgaps, although we note the existence of a strip mode near 6 \textrm{kHz} in the 1D Bloch band diagram that closes the bandgap of the macro-cell earlier than the corresponding point of the bulk band diagram. From observation of the mode shapes at the onset of the bandgaps of the coarse cell in Fig.~\ref{CoarseDenseCompare_SM}(f) and  Fig.~\ref{CoarseDenseCompare_SM}(h), we note again significant activation of flexural resonating units in the kerfing pattern, perhaps even more distinct here than in the case of the dense cells. %, and consistent between the 1D and 2D Bloch analyses.

\subsection*{\Large{Strategy to discriminate between strictly in-plane and out-of-plane modes}}

To distinguish between in-plane and out-of-plane modes of the band diagram, we conduct Bloch analysis on a purely planar model of the kerfed cell (with the same geometric and material parameters) using a 2D mesh of 4-node elements. Such analysis yields a band diagram whose modes are guaranteed to diplsay strictly in-plane character.
\begin{figure*}[!htpb]
	\centering
	\includegraphics[width=0.9\textwidth]{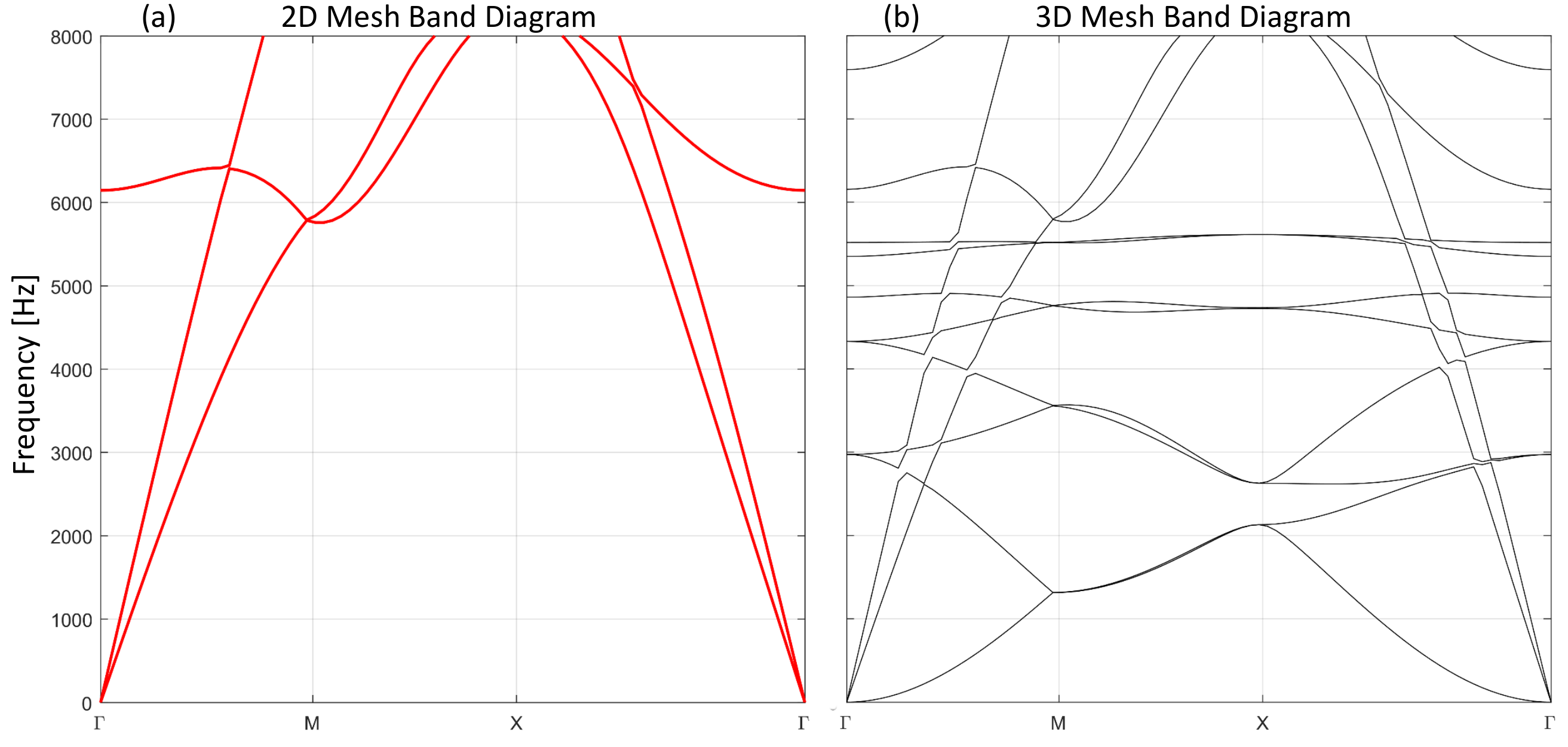}
	\caption{(a) Coarse cell band diagram computed with 2D Bloch analysis on a 2D FE mesh. This band diagram shows only in-plane modes due to the nature of the mesh and analysis conducted. (b) Coarse cell band diagram computed with 2D Bloch analysis on a 3D FE mesh which shows the full modal landscape of the unit cell, including in-plane modes.} \label{2D3DMesh_SM}
\end{figure*}

Fig.~\ref{2D3DMesh_SM}(a) shows the band diagram for a coarsely kerfed unit cell discretized with a 2D FE mesh, and we denote these in-plane modes in bold red to remain consistent with previously shown band diagrams. We compare this band diagram to the band diagram in Fig.~\ref{2D3DMesh_SM}(b) computed for the 3D model of a coarse unit cell, that captures all possible wave polarizations, i.e., both in-plane and out-of-plane modes. We observe that the red in-plane modes will directly overlie a subset of the modes obtained with the 3D analysis, deeming them as strictly in-plane modes. The remainder branches are therefore recognized as mode characterized by an out-of-plane component. This analysis allows us to immediately identify flexural bandgaps as frequency ranges that do not contain modes with out-of-plane character.

\begin{figure*}[!htbp]
	\centering
	\includegraphics[width=0.75\textwidth]{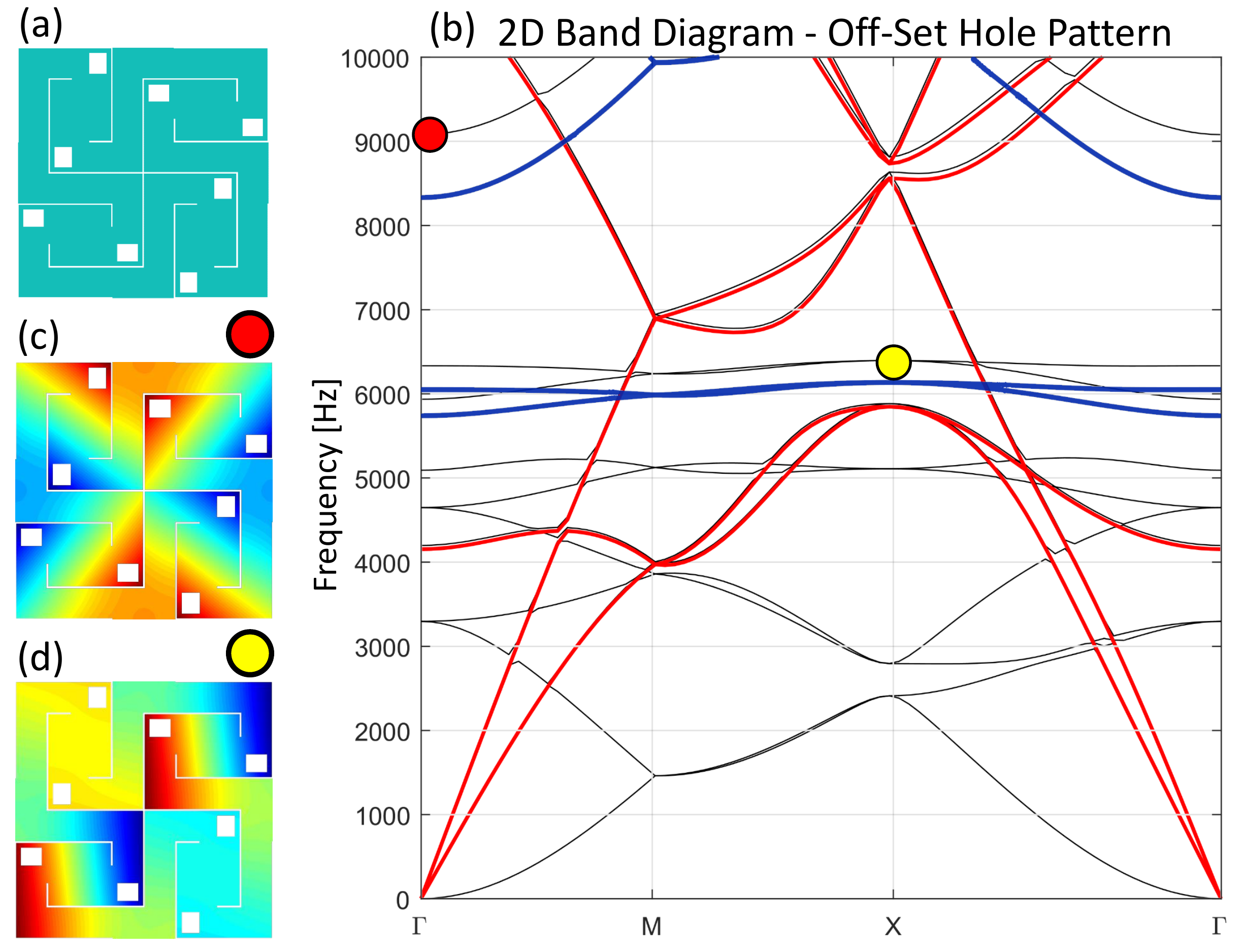}
	\caption{(a) Coarse unit cell with an off-set hole pattern. (b) 2D Bloch band diagram of the unit cell with off-set holes. Blue lines are modal branches from the band diagram of the unit cell without holes. The modal branches corresponding to the red and yellow dots show an increased frequency in the new band diagram. (c)-(d) Mode shapes at selected modes in the band structure which feature significant upshift in frequency caused by introduction of the hole pattern to the kerfed unit cell.} \label{OffSetHole_SM}
\end{figure*}

\subsection*{\Large{Evidence of resonant behavior via hole placement strategy}}

We can provide additional insight into the resonant nature of the bandgap-inducing mechanisms by investigating a set of modified kerf patterns featuring off-set micro-perforations. We take a coarsely kerfed cell and introduce small holes at the points shown in Fig.~\ref{OffSetHole_SM}(a) with the goal of modifying the band structure in an appreciable capacity.

Fig.~\ref{OffSetHole_SM}(b) shows the band diagram of the unit cell with micro-perforations, and we can immediately observe, by comparison with the modes from the band structure of the cell without holes (blue solid lines) that there is a significant upshift in the frequencies of certain modes. Particularly, if we look at the mode marked by the red point, we observe an upshift by several hundreds of Hz afforded by the introduction of the holes. We plot the mode shape corresponding to this spectral point in Fig.~\ref{OffSetHole_SM}(c). One can see that the holes have been deliberately placed in regions of the unit cell that experience significant deflection (dark red and dark blue in the color map), alternated by nodes of the mode shape (green regions) with little-to-no displacement. 

Let us consider the meandering beam in the interior of each a quarter cell as a structure undergoing flexural deformation, with the node of the mode shape acting as a clamp-like constraint and the regions undergoing large deflections as the free edges. Functionally, notwithstanding the differences in geometry, this can be modeled as a flexural resonator (conceptually analogous to a cantilever platelet), in which the stiffness is predominantly controlled by the elastic properties of the portion close to the root and the inertia by the mass density of the outer regions. Following this analogy, we can see that our hole placement scheme effectively removed material near the edges of the resonator, thus lowering its inertia. Further invoking a lumped-parameter model of the resonator, since the square of the natural frequency of a resonator is inversely proportional to its mass, the introduction of the holes is expected to cause an increase in resonant frequencies. Given that branches of the band diagram indeed migrate towards higher frequencies upon introduction of the micro-perforations, this suggests that the physics behind these modes could be pinpointed to the onset of these resonant mechanisms.

\end{document}